\documentclass[prl,english,reprint,nofootinbib,aps,superscriptaddress,showpacs,showkeys]{revtex4-1}

\usepackage{babel,calc,amsmath,amsthm,amssymb,graphicx,subfigure,xcolor,mathtools,
      longtable,braket,bm,extarrows}

\usepackage[T1]{fontenc}
\usepackage[unicode=true]{hyperref}
\hypersetup{
     colorlinks=true,       		
     linkcolor=blue,          	
     citecolor=red,            
     urlcolor=magenta,           	
}
\newtheorem*{theorem*}{Theorem}

\newtheorem*{corollary*}{Corollary}

\newtheorem*{lemma*}{Lemma}

\newtheorem*{proposition*}{Proposition}

\newtheorem*{conjecture*}{Conjecture}
\theoremstyle{definition}

\newtheorem*{definition*}{Definition}
\theoremstyle{remark}

\newtheorem*{remark*}{Remark}

\begin{document}

   \title{Identifying the $3$-qubit $W$ state with quantum uncertainty relation}

\author{Zhi-Jie Liu}
\affiliation{Theoretical Physics Division, Chern Institute of Mathematics and LPMC, Nankai University, Tianjin 300071, People's Republic of China}

\author{Hao-Nan Qiang}
\affiliation{Theoretical Physics Division, Chern Institute of Mathematics and LPMC, Nankai University, Tianjin 300071,
People's Republic of China}

\author{Jie Zhou}
\affiliation{College of Physics and Materials Science, Tianjin Normal University, Tianjin 300382, People's Republic of China}

\author{Mi Xie}
\affiliation{Department of Physics, School of Science, Tianjin University, Tianjin 300072,  People's Republic of China}

\author{Jing-Ling~Chen}
\email{chenjl@nankai.edu.cn}
\affiliation{Theoretical Physics Division, Chern Institute of Mathematics and LPMC, Nankai University, Tianjin 300071, People's Republic of China}

   \date{\today}
   \begin{abstract}
The $W$ state, a canonical representative of multipartite quantum
entanglement, plays a crucial role in quantum information science due to its
robust entanglement properties. Quantum uncertainty relations, on the other
hand, are a fundamental cornerstone of quantum mechanics. This paper
introduces a novel approach to Identifying tripartite $W$ states by leveraging
tripartite quantum uncertainty relations. By employing a specific set of
non-commuting observables, we formulate an uncertainty-based criterion for
identifying $W$ states and rigorously demonstrate its generality in
distinguishing them from other tripartite entangled states, such as the Greenberger-Horne-Zeilinger
state. Our approach bypasses the need for complete quantum state tomography,
as it requires only the verification of a set of uncertainty inequalities for
efficient $W$-state identification. This work provides a new theoretical tool
for identifying multipartite entangled states and underscores the significant
role of quantum uncertainty relations in entanglement characterization.

   \end{abstract}
   \maketitle

\section{Introduction}

Quantum entanglement \cite{Quantum entanglement2009}, as one of the most
distinctive features of quantum mechanics, plays a central role in quantum
information processing, quantum computing, quantum communication, and related
fields. Among multiparticle entangled states, the Greenberger-Horne-Zeilinger
state (GHZ state) \cite{GHZ1989} and the $W$ state \cite{WD2000} represent two
canonical classes of tripartite entanglement in $3$-qubit systems, yet they
possess fundamentally different characteristics. The GHZ state exhibits global
entanglement characteristics but undergoes complete decoherence under
single-particle measurements. In contrast, the $W$ state demonstrates
remarkable robustness: even upon the loss of one particle, the remaining
two-particle system retains bipartite entanglement. This unique property
affords $W$ states distinct advantages in applications such as quantum
networks and fault-tolerant quantum computing \cite{WD2000}.

Efficiently and reliably distinguishing between these two types of entangled
states, nevertheless, remains a critical challenge. Currently, the
identification of GHZ states relies on several well-established methods,
primarily including Bell inequality-based criteria \cite{ZC2004,XYF2021},
quantum state tomography \cite{KJR2005}, and entanglement witnesses
\cite{LHZ2024}. In contrast, the criterions for $W$ states remain rather
limited. In recent years, some studies have attempted to identify $W$ states
using entanglement witnesses \cite{MB2004}, eigenvalue-based quantum state
verification \cite{DB2024}, or quantum state tomography \cite{HM2005}. Such
methods, however, are constrained by either substantial experimental
complexity or limited universality. Therefore, developing an efficient and
universal $W$ state identification method is of significant importance.

The quantum uncertainty relation \cite{HSB1927} stands as one of the
fundamental principles of quantum mechanics, and its applications in many-body
systems have seen considerable advances in recent years. Traditionally,
uncertainty relations focused on pairs of single operators or scalar conserved
quantities \cite{ACF1996,HFH2003,OG2003,PB2007}. With the rapid development of
quantum technologies in recent years, research on uncertainty relations has
expanded from single operator pairs to multi-component systems
\cite{HHQ2019,ZXC2019,YLM2023,XZ2023,PJC2017,VVD2018,SK2014,QCS2016,WM2017,XBL2024,LZJ2025,LZJ20252}%
. These advances have provided a theoretical foundation for quantum precision
measurement, quantum communication, and quantum computing.

This paper proposes a novel approach for identifying three-qubit $W$ states
based on triple components uncertainty relations. Using spin-$\frac{1}{2}$
particles in an $XXZ$ Heisenberg model \cite{LJ2012} as our research system,
we analyze both additive and multiplicative uncertainty relations of spin
correlation operators. Our finding reveals that when these uncertainty
relations simultaneously satisfy their equality conditions, the quantum state
of the system must be a $W$ state. This criterion not only possesses rigorous
universality but can also effectively distinguish $W$ states from other
three-body entangled states. Overall, our method establishes for the first
time a direct connection between triple components uncertainty relations and
$W$ state identification, thereby filling a critical gap in efficient
discrimination tools for $W$ states.

\section{Main Results}

The Heisenberg model is a fundamental theoretical framework for studying
magnetic properties and quantum many-body systems. In this work, we
investigate the $XXZ$ Heisenberg model for a system composed of three
distinguishable spin-$\frac{1}{2}$ particles. The Hamiltonian is given by%
\begin{equation}
H=J\left(  H_{12}+H_{23}+H_{31}\right)  ,
\end{equation}
where%
\[
\left\{
\begin{array}
[c]{c}%
H_{12}=\hat{s}_{1x}\hat{s}_{2x}+\hat{s}_{1y}\hat{s}_{2y}+\gamma\hat{s}%
_{1z}\hat{s}_{2z},\\
H_{23}=\hat{s}_{2x}\hat{s}_{3x}+\hat{s}_{2y}\hat{s}_{3y}+\gamma\hat{s}%
_{2z}\hat{s}_{3z},\\
H_{31}=\hat{s}_{3x}\hat{s}_{1x}+\hat{s}_{3y}\hat{s}_{1y}+\gamma\hat{s}%
_{3z}\hat{s}_{1z},
\end{array}
\right.
\]
here, $J$ and $\gamma$ are constants. We analyze the uncertainty relations for
$H_{12}$, $H_{23}$, and $H_{31}$. For spin angular momentum, the commutation
relation is%
\[
\left[  \hat{s}_{u},\hat{s}_{v}\right]  =\mathrm{i}\hbar\epsilon_{uvw}\hat
{s}_{w},
\]
where $\epsilon_{uvw}$ is the Levi-Civita symbol, and $\epsilon_{xyz}%
=\epsilon_{yzx}=\epsilon_{zxy}=+1$, $\epsilon_{zyx}=\epsilon_{yxz}%
=\epsilon_{xzy}=-1$. We then compute the commutators%
\begin{align*}
&  \left[  H_{12},H_{23}\right] \\
=  &  -\mathrm{i}\hbar\left[  \vec{s}_{1}\cdot\left(  \vec{s}_{2}\times\vec
{s}_{3}\right)  \right. \\
&  +\left(  \gamma-1\right)  \left(  \hat{s}_{1x}\hat{s}_{2y}\hat{s}_{3z}%
-\hat{s}_{1y}\hat{s}_{2x}\hat{s}_{3z}\right. \\
&  +\left.  \left.  -\hat{s}_{1z}\hat{s}_{2y}\hat{s}_{3x}+\hat{s}_{1z}\hat
{s}_{2x}\hat{s}_{3y}\right)  \right]  ,
\end{align*}
here, $\vec{s}_{1}$, $\vec{s}_{2}$, and $\vec{s}_{3}$ represent the spin
angular momentum, and $\vec{s}_{i}=\left(  \hat{s}_{ix},\hat{s}_{iy},\hat
{s}_{iz}\right)  $. Similarly, the following expressions are derived%
\begin{align*}
&  \left[  H_{23},H_{31}\right] \\
=  &  -\mathrm{i}\hbar\left[  \vec{s}_{2}\cdot\left(  \vec{s}_{3}\times\vec
{s}_{1}\right)  \right. \\
&  +\left(  \gamma-1\right)  \left(  \hat{s}_{1z}\hat{s}_{2x}\hat{s}_{3y}%
-\hat{s}_{1z}\hat{s}_{2y}\hat{s}_{3x}\right. \\
&  +\left.  \left.  -\hat{s}_{1x}\hat{s}_{2z}\hat{s}_{3y}+\hat{s}_{1y}\hat
{s}_{2z}\hat{s}_{3x}\right)  \right]  ,
\end{align*}
and%
\begin{align*}
&  \left[  H_{31},H_{12}\right] \\
=  &  -\mathrm{i}\hbar\left[  \vec{s}_{3}\cdot\left(  \vec{s}_{1}\times\vec
{s}_{2}\right)  \right. \\
&  +\left(  \gamma-1\right)  \left(  \hat{s}_{1y}\hat{s}_{2z}\hat{s}_{3x}%
-\hat{s}_{1x}\hat{s}_{2z}\hat{s}_{3y}\right. \\
&  +\left.  \left.  -\hat{s}_{1y}\hat{s}_{2x}\hat{s}_{3z}+\hat{s}_{1x}\hat
{s}_{2y}\hat{s}_{3z}\right)  \right]  .
\end{align*}
Then we analyze both the additive and multiplicative uncertainty relations for
$H_{12}$, $H_{23}$, and $H_{31}$.

\subsection{Additive Uncertainty Relation}

We first consider the additive uncertainty relations. According to the
Heisenberg uncertainty principle
\begin{equation}
\triangle A\triangle B\geqslant\frac{1}{2}\left\vert \left\langle \left[
A,B\right]  \right\rangle \right\vert , \label{HSB}%
\end{equation}
where for any observable measurements $A$ and $B$, $\triangle A=\sqrt
{\left\langle A^{2}\right\rangle -\left\langle A\right\rangle ^{2}}$,
$\triangle B=\sqrt{\left\langle B^{2}\right\rangle -\left\langle
B\right\rangle ^{2}}$, and the angle brackets $\left\langle \left.  {}\right.
\right\rangle $ denote the expectation of an operator with respect to a given
state $\rho$. The commutation relation $\left[  A,B\right]  =AB-BA$. This
leads to%
\begin{equation}
\left\{
\begin{array}
[c]{c}%
\triangle H_{12}\triangle H_{23}\geqslant\frac{1}{2}\mathcal{S}_{1},\\
\triangle H_{23}\triangle H_{31}\geqslant\frac{1}{2}\mathcal{S}_{2},\\
\triangle H_{31}\triangle H_{12}\geqslant\frac{1}{2}\mathcal{S}_{3},
\end{array}
\right.  \label{s0}%
\end{equation}
where%
\begin{align*}
\mathcal{S}_{1}  &  =\hbar\left\vert \left\langle \vec{s}_{1}\cdot\left(
\vec{s}_{2}\times\vec{s}_{3}\right)  \right\rangle \right\vert \\
&  +\hbar\left(  \gamma-1\right)  \left\vert \left\langle \left(  \hat{s}%
_{1x}\hat{s}_{2y}\hat{s}_{3z}-\hat{s}_{1y}\hat{s}_{2x}\hat{s}_{3z}\right.
\right.  \right. \\
&  \left.  \left.  \left.  -\hat{s}_{1z}\hat{s}_{2y}\hat{s}_{3x}+\hat{s}%
_{1z}\hat{s}_{2x}\hat{s}_{3y}\right)  \right\rangle \right\vert ,
\end{align*}%
\begin{align*}
\mathcal{S}_{2}  &  =\hbar\left\vert \left\langle \vec{s}_{2}\cdot\left(
\vec{s}_{3}\times\vec{s}_{1}\right)  \right\rangle \right\vert \\
&  +\hbar\left(  \gamma-1\right)  \left\vert \left\langle \left(  \hat{s}%
_{1z}\hat{s}_{2x}\hat{s}_{3y}-\hat{s}_{1z}\hat{s}_{2y}\hat{s}_{3x}\right.
\right.  \right. \\
&  \left.  \left.  \left.  -\hat{s}_{1x}\hat{s}_{2z}\hat{s}_{3y}+\hat{s}%
_{1y}\hat{s}_{2z}\hat{s}_{3x}\right)  \right\rangle \right\vert ,
\end{align*}
and%
\begin{align*}
\mathcal{S}_{3}  &  =\hbar\left\vert \left\langle \vec{s}_{3}\cdot\left(
\vec{s}_{1}\times\vec{s}_{2}\right)  \right\rangle \right\vert \\
&  +\hbar\left(  \gamma-1\right)  \left\vert \left\langle \left(  \hat{s}%
_{1y}\hat{s}_{2z}\hat{s}_{3x}-\hat{s}_{1x}\hat{s}_{2z}\hat{s}_{3y}\right.
\right.  \right. \\
&  \left.  \left.  \left.  -\hat{s}_{1y}\hat{s}_{2x}\hat{s}_{3z}+\hat{s}%
_{1x}\hat{s}_{2y}\hat{s}_{3z}\right)  \right\rangle \right\vert .
\end{align*}
Moreover,%
\begin{align}
\left(  \triangle H_{12}\right)  ^{2}+\left(  \triangle H_{23}\right)  ^{2}
&  \geqslant2\triangle H_{12}\triangle H_{23}\nonumber\\
&  \geqslant\mathcal{S}_{1}, \label{s1}%
\end{align}
Similarly, we have%
\begin{equation}
\left(  \triangle H_{23}\right)  ^{2}+\left(  \triangle H_{31}\right)
^{2}\geqslant\mathcal{S}_{2}, \label{s2}%
\end{equation}
and%
\begin{equation}
\left(  \triangle H_{31}\right)  ^{2}+\left(  \triangle H_{12}\right)
^{2}\geqslant\mathcal{S}_{3}. \label{s3}%
\end{equation}
By summing Eqs. (\ref{s1}), (\ref{s2}), and (\ref{s3}), we arrive at%
\begin{align*}
\left(  \triangle H_{12}\right)  ^{2}  &  +\left(  \triangle H_{23}\right)
^{2}+\left(  \triangle H_{31}\right)  ^{2}\\
&  \geqslant\frac{1}{2}\left(  \mathcal{S}_{1}+\mathcal{S}_{2}+\mathcal{S}%
_{3}\right)  .
\end{align*}
However, the equality holds if and only if both sides vanish. Therefore, we
introduce a constant $\tau_{s}$ and reformulate the uncertainty relation as%
\begin{align*}
&  \left(  \triangle H_{12}\right)  ^{2}+\left(  \triangle H_{23}\right)
^{2}+\left(  \triangle H_{31}\right)  ^{2}\\
&  \geqslant\frac{\tau_{s}}{2}\left(  \mathcal{S}_{1}+\mathcal{S}%
_{2}+\mathcal{S}_{3}\right)  .
\end{align*}

Next, we study $\tau_{s}$. From the above equation, we can obtain
\begin{equation}
\tau_{s}\leqslant\frac{2\left[  \left(  \triangle H_{12}\right)  ^{2}+\left(
\triangle H_{23}\right)  ^{2}+\left(  \triangle H_{31}\right)  ^{2}\right]
}{\mathcal{S}_{1}+\mathcal{S}_{2}+\mathcal{S}_{3}}. \label{taus1}%
\end{equation}
Consider the most general $3$-qubit state%
\begin{align}
\left\vert \psi\right\rangle  &  =c_{1}\left\vert 000\right\rangle
+c_{2}\left\vert 001\right\rangle +c_{3}\left\vert 010\right\rangle
+c_{4}\left\vert 100\right\rangle \nonumber\\
&  +c_{5}\left\vert 011\right\rangle +c_{6}\left\vert 101\right\rangle
+c_{7}\left\vert 110\right\rangle +c_{8}\left\vert 111\right\rangle ,
\label{psi}%
\end{align}
where $\sum_{i=1}^{8}\left\vert c_{i}\right\vert ^{2}=1$.

After calculation, we simplify the result of Eq. (\ref{taus1}). The detailed
derivation is provided in the Appendix, where the denominator in Eq.
(\ref{taus1}) reduces to%
\begin{align*}
\mathcal{D}  &  \mathcal{\equiv S}_{1}+\mathcal{S}_{2}+\mathcal{S}_{3}\\
&  =\frac{\hbar^{4}}{4}\left(  \left\vert \mathcal{A}\right\vert +\left\vert
\mathcal{B}\right\vert +\left\vert \mathcal{C}\right\vert \right) \\
&  \geqslant\frac{\hbar^{4}}{4}\left\vert \mathcal{A}+\mathcal{B}%
+\mathcal{C}\right\vert .
\end{align*}
where%
\begin{align*}
\mathcal{A}  &  \equiv\mathrm{i}m+\mathrm{i}\left(  \gamma-1\right)  \left[
\left(  c_{2}^{\ast}-c_{4}^{\ast}\right)  c_{3}-c_{3}^{\ast}c_{2}\right. \\
&  \left.  +c_{3}^{\ast}c_{4}+\left(  c_{7}^{\ast}-c_{5}^{\ast}\right)
c_{6}-c_{6}^{\ast}c_{7}+c_{6}^{\ast}c_{5}\right]  ,
\end{align*}%
\begin{align*}
\mathcal{B}  &  \equiv\mathrm{i}m+\mathrm{i}\left(  \gamma-1\right)  \left[
\left(  c_{4}^{\ast}-c_{3}^{\ast}\right)  c_{2}-c_{2}^{\ast}c_{4}\right. \\
&  \left.  +c_{2}^{\ast}c_{3}+\left(  c_{5}^{\ast}-c_{6}^{\ast}\right)
c_{7}-c_{7}^{\ast}c_{5}+c_{7}^{\ast}c_{6}\right]  ,
\end{align*}%
\begin{align*}
\mathcal{C}  &  \equiv\mathrm{i}m+\mathrm{i}\left(  \gamma-1\right)  \left[
\left(  c_{3}^{\ast}-c_{2}^{\ast}\right)  c_{4}-c_{4}^{\ast}c_{3}\right. \\
&  \left.  +c_{4}^{\ast}c_{2}+\left(  c_{6}^{\ast}-c_{7}^{\ast}\right)
c_{5}-c_{5}^{\ast}c_{6}+c_{5}^{\ast}c_{7}\right]  ,
\end{align*}
and%
\begin{align*}
m  &  =\left(  c_{4}^{\ast}-c_{3}^{\ast}\right)  c_{2}+\left(  c_{2}^{\ast
}-c_{4}^{\ast}\right)  c_{3}+\left(  c_{3}^{\ast}-c_{2}^{\ast}\right)  c_{4}\\
&  +\left(  c_{6}^{\ast}-c_{7}^{\ast}\right)  c_{5}+\left(  c_{7}^{\ast}%
-c_{5}^{\ast}\right)  c_{6}+\left(  c_{5}^{\ast}-c_{6}^{\ast}\right)  c_{7}.
\end{align*}
The calculation results in%
\[
\mathcal{A}+\mathcal{B}+\mathcal{C}=\left(  2\gamma+1\right)  \left\langle
\psi\right\vert \mathcal{P}\left\vert \psi\right\rangle ,
\]
with%
\[
\mathcal{P=}\left(
\begin{array}
[c]{cccc}%
0 &  &  & \\
& P &  & \\
&  & P^{\top} & \\
&  &  & 0
\end{array}
\right)  ,P=\left(
\begin{array}
[c]{ccc}%
0 & \mathrm{i} & -\mathrm{i}\\
-\mathrm{i} & 0 & \mathrm{i}\\
\mathrm{i} & -\mathrm{i} & 0
\end{array}
\right)  .
\]
Assuming the eigenvalues of $P$ are $\lambda_{i}$ with corresponding
eigenstates $\left\vert \psi_{i}\right\rangle $, solving the eigenvalue
equation yields the eigenvalues, we obtain $\lambda_{5}=\lambda_{6}%
=\lambda_{7}=\lambda_{8}=0$, $\lambda_{3}=\lambda_{4}=\sqrt{3}$, and
$\lambda_{1}=\lambda_{2}=-\sqrt{3}$. Let the quantum state be expressed as
$\left\vert \psi\right\rangle =\sum_{i}\sqrt{p_{i}}\mathrm{e}^{\mathrm{i}%
\varphi_{i}}\left\vert \psi_{i}\right\rangle $, where the eight eigenstates
are denoted by%
\begin{equation}
\left\{
\begin{array}
[c]{l}%
\left\vert \psi_{1}\right\rangle =\frac{\mathrm{e}^{\mathrm{i}\frac{4\pi}{3}}%
}{\sqrt{3}}\left\vert 011\right\rangle +\frac{\mathrm{e}^{\mathrm{i}\frac
{2\pi}{3}}}{\sqrt{3}}\left\vert 101\right\rangle +\frac{1}{\sqrt{3}}\left\vert
110\right\rangle ,\\
\left\vert \psi_{2}\right\rangle =\frac{\mathrm{e}^{\mathrm{i}\frac{2\pi}{3}}%
}{\sqrt{3}}\left\vert 001\right\rangle +\frac{\mathrm{e}^{\mathrm{i}\frac
{4\pi}{3}}}{\sqrt{3}}\left\vert 010\right\rangle +\frac{1}{\sqrt{3}}\left\vert
100\right\rangle ,\\
\left\vert \psi_{3}\right\rangle =\frac{\mathrm{e}^{\mathrm{i}\frac{2\pi}{3}}%
}{\sqrt{3}}\left\vert 011\right\rangle +\frac{\mathrm{e}^{\mathrm{i}\frac
{4\pi}{3}}}{\sqrt{3}}\left\vert 101\right\rangle +\frac{1}{\sqrt{3}}\left\vert
110\right\rangle ,\\
\left\vert \psi_{4}\right\rangle =\frac{\mathrm{e}^{\mathrm{i}\frac{4\pi}{3}}%
}{\sqrt{3}}\left\vert 001\right\rangle +\frac{\mathrm{e}^{\mathrm{i}\frac
{2\pi}{3}}}{\sqrt{3}}\left\vert 010\right\rangle +\frac{1}{\sqrt{3}}\left\vert
100\right\rangle ,\\
\left\vert \psi_{5}\right\rangle =\left\vert 111\right\rangle ,\\
\left\vert \psi_{6}\right\rangle =\frac{1}{\sqrt{3}}\left\vert
011\right\rangle +\frac{1}{\sqrt{3}}\left\vert 101\right\rangle +\frac
{1}{\sqrt{3}}\left\vert 110\right\rangle ,\\
\left\vert \psi_{7}\right\rangle =\frac{1}{\sqrt{3}}\left\vert
001\right\rangle +\frac{1}{\sqrt{3}}\left\vert 010\right\rangle +\frac
{1}{\sqrt{3}}\left\vert 100\right\rangle ,\\
\left\vert \psi_{8}\right\rangle =\left\vert 000\right\rangle .
\end{array}
\right.  \label{psii}%
\end{equation}
Then%
\[
\mathcal{D}\geqslant\frac{\sqrt{3}\hbar^{4}}{4}\left\vert \left(
2\gamma+1\right)  \left(  -p_{1}-p_{2}+p_{3}+p_{4}\right)  \right\vert .
\]

Proceeding to the numerator $\mathcal{N}$, after simplification we have%
\begin{align*}
\mathcal{N}=  &  2\left(  \frac{\hbar}{2}\right)  ^{4}\left\{  2\left(
\gamma+1\right)  ^{2}\left(  f_{1}+g_{1}+h_{1}\right)  \right. \\
&  +2\left(  \gamma-1\right)  ^{2}\left(  f_{2}+g_{2}+h_{2}\right) \\
&  -\left[  \left(  \gamma+1\right)  g_{1}+\left(  \gamma-1\right)
g_{2}\right]  ^{2}\\
&  -\left[  \left(  \gamma+1\right)  f_{1}+\left(  \gamma-1\right)
f_{2}\right]  ^{2}\\
&  -\left.  \left[  \left(  \gamma+1\right)  h_{1}+\left(  \gamma-1\right)
h_{2}\right]  ^{2}\right\}  ,
\end{align*}
where%
\begin{equation}
\left\{
\begin{array}
[c]{c}%
\left(  c_{4}^{\ast}-c_{3}^{\ast}\right)  \left(  c_{4}-c_{3}\right)  +\left(
c_{5}^{\ast}-c_{6}^{\ast}\right)  \left(  c_{5}-c_{6}\right)  =f_{1},\\
\left(  c_{4}^{\ast}+c_{3}^{\ast}\right)  \left(  c_{4}+c_{3}\right)  +\left(
c_{5}^{\ast}+c_{6}^{\ast}\right)  \left(  c_{5}+c_{6}\right)  =f_{2},\\
\left(  c_{4}^{\ast}-c_{2}^{\ast}\right)  \left(  c_{4}-c_{2}\right)  +\left(
c_{5}^{\ast}-c_{7}^{\ast}\right)  \left(  c_{5}-c_{7}\right)  =g_{1},\\
\left(  c_{4}^{\ast}+c_{2}^{\ast}\right)  \left(  c_{4}+c_{2}\right)  +\left(
c_{5}^{\ast}+c_{7}^{\ast}\right)  \left(  c_{5}+c_{7}\right)  =g_{2},\\
\left(  c_{3}^{\ast}-c_{2}^{\ast}\right)  \left(  c_{3}-c_{2}\right)  +\left(
c_{6}^{\ast}-c_{7}^{\ast}\right)  \left(  c_{6}-c_{7}\right)  =h_{1},\\
\left(  c_{3}^{\ast}+c_{2}^{\ast}\right)  \left(  c_{3}+c_{2}\right)  +\left(
c_{6}^{\ast}+c_{7}^{\ast}\right)  \left(  c_{6}+c_{7}\right)  =h_{2}.
\end{array}
\right.
\end{equation}
Since $f^{2}+g^{2}+h^{2}\geqslant fg+fh+hg$, where the equality holds if and
only if $f=g=h$, it follows that%
\begin{align*}
\mathcal{N}\leqslant &  \mathcal{N}+\frac{2}{3}\left\{  \sum_{d=f,g,h}\left[
\left(  \gamma+1\right)  d_{1}+\left(  \gamma-1\right)  d_{2}\right]
^{2}\right. \\
&  \left.  -\left[  \left(  \gamma+1\right)  f_{1}+\left(  \gamma-1\right)
f_{2}\right]  \left[  \left(  \gamma+1\right)  g_{1}+\left(  \gamma-1\right)
g_{2}\right]  \right\} \\
&  -\left[  \left(  \gamma+1\right)  f_{1}+\left(  \gamma-1\right)
f_{2}\right]  \left[  \left(  \gamma+1\right)  h_{1}+\left(  \gamma-1\right)
h_{2}\right] \\
&  -\left[  \left(  \gamma+1\right)  g_{1}+\left(  \gamma-1\right)
g_{2}\right]  \left[  \left(  \gamma+1\right)  h_{1}+\left(  \gamma-1\right)
h_{2}\right] \\
=  &  2\left(  \frac{\hbar}{2}\right)  ^{4}\left\{  2\left(  \gamma+1\right)
^{2}\left\langle \psi\right\vert \mathcal{F}\left\vert \psi\right\rangle
+2\left(  \gamma-1\right)  ^{2}\left\langle \psi\right\vert \mathcal{G}%
\left\vert \psi\right\rangle \right. \\
&  \left.  -\frac{1}{3}\left[  \left(  \gamma+1\right)  \left\langle
\psi\right\vert \mathcal{F}\left\vert \psi\right\rangle +\left(
\gamma-1\right)  \left\langle \psi\right\vert \mathcal{G}\left\vert
\psi\right\rangle \right]  ^{2}\right\}  ,
\end{align*}
where%
\[
\left\{
\begin{array}
[c]{c}%
\left\langle \psi\right\vert \mathcal{F}\left\vert \psi\right\rangle \equiv
f_{1}+g_{1}+h_{1},\\
\left\langle \psi\right\vert \mathcal{G}\left\vert \psi\right\rangle \equiv
f_{2}+g_{2}+h_{2},
\end{array}
\right.
\]
and%
\[
\mathcal{F=}\left(
\begin{array}
[c]{cccc}%
0 &  &  & \\
& P^{2} &  & \\
&  & P^{2} & \\
&  &  & 0
\end{array}
\right)  ,
\]%
\[
\mathcal{G=}\left(
\begin{array}
[c]{cccc}%
0 &  &  & \\
& G &  & \\
&  & G & \\
&  &  & 0
\end{array}
\right)  ,
\]
with%
\[
G=\left(
\begin{array}
[c]{ccc}%
2 & 1 & 1\\
1 & 2 & 1\\
1 & 1 & 2
\end{array}
\right)  .
\]
Since $\mathcal{F}$ and $\mathcal{G}$ commute, they share a common set of
eigenstates given by Eq. (\ref{psii}). Let $\alpha_{i}$ and $\beta_{j}$ denote
the eigenvalues of $\mathcal{F}$ and $\mathcal{G}$, respectively. The
eigenvalue spectra are $\alpha_{1}=\alpha_{2}=\alpha_{3}=\alpha_{4}=3$,
$\alpha_{5}=\alpha_{6}=\alpha_{7}=\alpha_{8}=0$, $\beta_{1}=\beta_{2}%
=\beta_{3}=\beta_{4}=1$, $\beta_{6}=\beta_{7}=4$, and $\beta_{5}=\beta_{8}=0$.
Consequently,
\begin{align*}
\mathcal{N}  &  \leqslant\frac{\hbar^{4}}{8}\left\{  8\left(  1+\gamma
+\gamma^{2}\right)  \sum_{i=1}^{4}p_{i}+8\left(  1-2\gamma+\gamma^{2}\right)
\sum_{j=6}^{7}p_{j}\right. \\
&  \left.  -\frac{1}{3}\left[  2\left(  2\gamma+1\right)  \sum_{i=1}^{4}%
p_{i}+4\left(  \gamma-1\right)  \sum_{j=6}^{7}p_{j}\right]  ^{2}\right\}  .
\end{align*}
We define%
\[
\left\{
\begin{array}
[c]{c}%
\dfrac{p_{3}+p_{4}}{\sum_{i=1}^{7}p_{i}}=u_{1},\\
\dfrac{p_{1}+p_{2}}{\sum_{i=1}^{7}p_{i}}=u_{2}.\\
\dfrac{p_{6}+p_{7}}{\sum_{i=1}^{7}p_{i}}=u_{3}.
\end{array}
\right.
\]
This leads to
\begin{align*}
\tau_{s}  &  \leqslant\frac{\mathcal{N}}{\mathcal{D}}\\
&  =\frac{8\left(  1+\gamma+\gamma^{2}\right)  \left(  u_{2}+u_{1}\right)
+8\left(  1-2\gamma+\gamma^{2}\right)  u_{3}}{2\sqrt{3}\left\vert \left(
2\gamma+1\right)  \left(  -u_{2}+u_{1}\right)  \right\vert }\\
&  -\frac{\frac{1}{3}\sum_{i=1}^{7}p_{i}\left[  2\left(  2\gamma+1\right)
\left(  u_{2}+u_{1}\right)  +4\left(  \gamma-1\right)  u_{3}\right]  ^{2}%
}{2\sqrt{3}\left\vert \left(  2\gamma+1\right)  \left(  -u_{2}+u_{1}\right)
\right\vert }.
\end{align*}
Since $0\leqslant\sum_{i=1}^{7}p_{i}\leqslant1$, the ratio $\frac{\mathcal{N}%
}{\mathcal{D}}$ is minimized when $\sum_{i=1}^{7}p_{i}=1$. Under this
condition, we have%
\begin{align*}
\tau_{s}\leqslant &  \frac{8\left(  1+\gamma^{2}\right)  +8\gamma\left(
u_{1}+u_{2}-2u_{3}\allowbreak\right)  }{2\sqrt{3}\left\vert \left(
2\gamma+1\right)  \left(  -u_{2}+u_{1}\right)  \right\vert }\\
&  -\frac{\frac{1}{3}\left[  2\left(  u_{1}+u_{2}-2u_{3}\right)
+4\gamma\right]  ^{2}}{2\sqrt{3}\left\vert \left(  2\gamma+1\right)  \left(
-u_{2}+u_{1}\right)  \right\vert }\\
=  &  \frac{12\left(  1+\gamma^{2}\right)  +12\gamma\left(  u_{1}+u_{2}%
-2u_{3}\allowbreak\right)  }{3\sqrt{3}\left\vert \left(  2\gamma+1\right)
\left(  -u_{2}+u_{1}\right)  \right\vert }\\
&  -\frac{2\left[  \left(  u_{1}+u_{2}-2u_{3}\right)  +2\gamma\right]  ^{2}%
}{3\sqrt{3}\left\vert \left(  2\gamma+1\right)  \left(  -u_{2}+u_{1}\right)
\right\vert }\\
=  &  \frac{\left(  2\gamma+1\right)  ^{2}+6\left(  2\gamma+1\right)  u_{3}%
}{3\sqrt{3}\left\vert \left(  2\gamma+1\right)  \left(  u_{1}-u_{2}\right)
\right\vert }\\
+  &  \frac{13-2\left(  1-3u_{3}\right)  \left(  2-3u_{3}\right)  }{3\sqrt
{3}\left\vert \left(  2\gamma+1\right)  \left(  u_{1}-u_{2}\right)
\right\vert }\\
=  &  \frac{\left\vert 2\gamma+1\right\vert \mp6u_{3}+\frac{13-2\left(
1-3u_{3}\right)  \left(  2-3u_{3}\right)  }{\left\vert 2\gamma+1\right\vert }%
}{3\sqrt{3}\left\vert \left(  u_{1}-u_{2}\right)  \right\vert }.
\end{align*}
Since $\left\vert 2\gamma+1\right\vert +\frac{13-2\left(  1-3u_{3}\right)
\left(  2-3u_{3}\right)  }{\left\vert 2\gamma+1\right\vert }$ $\geqslant
2\sqrt{13-2\left(  1-3u_{3}\right)  \left(  2-3u_{3}\right)  }$, where
equality holds if and only if $\left\vert 2\gamma+1\right\vert =\frac
{13-2\left(  1-3u_{3}\right)  \left(  2-3u_{3}\right)  }{\left\vert
2\gamma+1\right\vert }$, we derive the bound%
\begin{align*}
\tau_{s}  &  \leqslant\frac{\mp6u_{3}+2\sqrt{13-2\left(  1-3u_{3}\right)
\left(  2-3u_{3}\right)  }}{3\sqrt{3}\left\vert \left(  u_{1}-u_{2}\right)
\right\vert }\\
&  =\frac{2}{\sqrt{3}}\frac{\mp u_{3}+\sqrt{1+2u_{3}-2u_{3}^{2}}}{\left\vert
\left(  u_{1}-u_{2}\right)  \right\vert }.
\end{align*}
Since $0\leqslant\left\vert u_{1}-u_{2}\right\vert \leqslant1-u_{3}$, the
upper bound for $\tau_{s}$ is minimized when $\left\vert u_{1}-u_{2}%
\right\vert =1-u_{3}$, yielding%
\[
\tau_{s}\leqslant\frac{2}{\sqrt{3}}\frac{\mp u_{3}+\sqrt{1+2u_{3}-2u_{3}^{2}}%
}{1-u_{3}}.
\]
We analyze the function $\mathcal{T}\left(  u_{3}\right)  =\frac{\mp
u_{3}+\sqrt{1+2u_{3}-2u_{3}^{2}}}{1-u_{3}}$ by taking its derivative, the
result shows%
\[
\frac{d\mathcal{T}}{du_{3}}=\frac{2-u_{3}\pm\sqrt{1+2u_{3}-2u_{3}^{2}}%
}{\left(  1-u_{3}\right)  ^{2}\sqrt{1+2u_{3}-2u_{3}^{2}}}.
\]
Given the parameter range $0\leqslant u_{3}\leqslant1$, the non-negative
derivative condition $\frac{d\mathcal{T}}{du_{3}}\geqslant0$ holds throughout
this interval, ensuring that $\mathcal{T}$ attains its minimum value
$\mathcal{T}_{\min}=1$ at the boundary point $u_{3}=0$. Consequently,
$\tau_{s}\leqslant\frac{2}{\sqrt{3}}$, where the equality condition requires
both $u_{3}=0$ and $\left\vert u_{1}-u_{2}\right\vert =1$, leading to two
possible configurations with either $\left(  u_{1},u_{2}\right)  =\left(
1,0\right)  $ or $\left(  0,1\right)  $. Under the $u_{3}=0$ condition, the
constraint $\left\vert 2\gamma+1\right\vert =3$ gives two solutions for the
anisotropy parameter, $\gamma=1$ and $\gamma=-2$, corresponding to distinct
physical regimes.

\begin{enumerate}
\item $\gamma=1$

\begin{enumerate}
\item $u_{1}=0$, $u_{2}=1$%
\begin{equation}
\left\vert \psi\right\rangle _{c_{1}}=\sqrt{1-p_{1}}\mathrm{e}^{\mathrm{i}%
\varphi_{1}}\left\vert \psi_{1}\right\rangle +\sqrt{p_{1}}\mathrm{e}%
^{\mathrm{i}\varphi_{2}}\left\vert \psi_{2}\right\rangle .
\end{equation}

\item $u_{1}=1$, $u_{2}=0$%
\begin{equation}
\left\vert \psi\right\rangle _{c_{2}}=\sqrt{1-p_{3}}\mathrm{e}^{\mathrm{i}%
\varphi_{3}}\left\vert \psi_{3}\right\rangle +\sqrt{p_{3}}\mathrm{e}%
^{\mathrm{i}\varphi_{4}}\left\vert \psi_{4}\right\rangle .
\end{equation}

\end{enumerate}

\item $\gamma=-2$

\begin{enumerate}
\item $u_{1}=0$, $u_{2}=1$%
\begin{equation}
\left\vert \psi\right\rangle _{c_{3}}=\left\vert \psi\right\rangle _{c_{1}%
}=\sqrt{1-p_{1}}\mathrm{e}^{\mathrm{i}\varphi_{1}}\left\vert \psi
_{1}\right\rangle +\sqrt{p_{1}}\mathrm{e}^{\mathrm{i}\varphi_{2}}\left\vert
\psi_{2}\right\rangle .
\end{equation}

\item $u_{1}=1$, $u_{2}=0$%
\begin{equation}
\left\vert \psi\right\rangle _{c_{4}}=\left\vert \psi\right\rangle _{c_{2}%
}=\sqrt{1-p_{3}}\mathrm{e}^{\mathrm{i}\varphi_{3}}\left\vert \psi
_{3}\right\rangle +\sqrt{p_{3}}\mathrm{e}^{\mathrm{i}\varphi_{4}}\left\vert
\psi_{4}\right\rangle .
\end{equation}

\end{enumerate}
\end{enumerate}

In summary, when $\gamma=1$ or $-2$, $\tau_{s\max}=\frac{2}{\sqrt{3}}$, at
which point
\begin{equation}
\left\vert \psi\right\rangle _{c_{1}}=\sqrt{1-p_{1}}\left\vert \psi
_{1}\right\rangle +\sqrt{p_{1}}\mathrm{e}^{\mathrm{i}\varphi}\left\vert
\psi_{2}\right\rangle ,
\end{equation}
and%
\begin{equation}
\left\vert \psi\right\rangle _{c_{2}}=\sqrt{1-p_{3}}\left\vert \psi
_{3}\right\rangle +\sqrt{p_{3}}\mathrm{e}^{\mathrm{i}\varphi^{\prime}%
}\left\vert \psi_{4}\right\rangle ,
\end{equation}
here $p_{1}$, $p_{3}\in\left[  0,1\right]  $ and $\varphi$, $\varphi\prime
\in\left[  0,2\pi\right]  $.

It is worth noting that we set%
\[
\left\{
\begin{array}
[c]{l}%
\left\vert 0^{\prime}\right\rangle =\sqrt{p_{1}}\left\vert 0\right\rangle
-\sqrt{1-p_{1}}\mathrm{e}^{\mathrm{i}\left(  \frac{4\pi}{3}-\varphi\right)
}\left\vert 1\right\rangle ,\\
\left\vert 1^{\prime}\right\rangle =\sqrt{1-p_{1}}\left\vert 0\right\rangle
+\sqrt{p_{1}}\mathrm{e}^{\mathrm{i}\left(  \frac{4\pi}{3}-\varphi\right)
}\left\vert 1\right\rangle ,
\end{array}
\right.
\]
then%
\begin{align*}
&  \left\vert \psi\right\rangle _{c_{1}}^{\prime}\\
&  =\frac{\mathrm{e}^{\mathrm{i}\left(  \frac{2\pi}{3}+2\varphi\right)  }%
}{\sqrt{3}}\left(  \mathrm{e}^{\mathrm{i}\frac{2\pi}{3}}\left\vert 0^{\prime
}0^{\prime}1^{\prime}\right\rangle +\mathrm{e}^{\mathrm{i}\frac{4\pi}{3}%
}\left\vert 0^{\prime}1^{\prime}0^{\prime}\right\rangle +\left\vert 1^{\prime
}0^{\prime}0^{\prime}\right\rangle \right)  .
\end{align*}
Furthermore, by setting%
\[
\left\{
\begin{array}
[c]{c}%
\left\vert 0^{\prime\prime}\right\rangle =\sqrt{p_{3}}\left\vert
0\right\rangle -\sqrt{1-p_{3}}\mathrm{e}^{\mathrm{i}\left(  \frac{2\pi}%
{3}-\varphi^{\prime}\right)  }\left\vert 1\right\rangle ,\\
\left\vert 1^{\prime\prime}\right\rangle =\sqrt{1-p_{3}}\left\vert
0\right\rangle +\sqrt{p_{3}}\mathrm{e}^{\mathrm{i}\left(  \frac{2\pi}%
{3}-\varphi^{\prime}\right)  }\left\vert 1\right\rangle ,
\end{array}
\right.
\]
we obtain%
\begin{align*}
&  \left\vert \psi\right\rangle _{c_{2}}^{\prime}\\
&  =\frac{\mathrm{e}^{\mathrm{i}\left(  \frac{4\pi}{3}+2\varphi^{\prime
}\right)  }}{\sqrt{3}}\left(  \mathrm{e}^{\mathrm{i}\frac{4\pi}{3}}\left\vert
0^{\prime\prime}0^{\prime\prime}1^{\prime\prime}\right\rangle +\mathrm{e}%
^{\mathrm{i}\frac{2\pi}{3}}\left\vert 0^{\prime\prime}1^{\prime\prime
}0^{\prime\prime}\right\rangle +\left\vert 1^{\prime\prime}0^{\prime\prime
}0^{\prime\prime}\right\rangle \right)  .
\end{align*}
It can be observed that both $\left\vert \psi\right\rangle _{c_{1}}^{\prime}$
and $\left\vert \psi\right\rangle _{c_{2}}^{\prime}$ are W states. This means
that in the $XXZ$ Heisenberg model, when the uncertainty relations for the
three physical quantities $H_{12}$, $H_{23}$, and $H_{31}$ saturate (i.e.,
reach equality), the critical state is a $W$ state. Through these uncertainty
relations, we have derived a criterion for identifying $W$ states.

\subsection{Multiplicative Uncertainty Relation}

We consider the multiplicative uncertainty relation. From Eq. (\ref{s0}), we
obtain%
\[
\triangle H_{12}\triangle H_{23}\triangle H_{31}\geqslant\sqrt{\frac{1}%
{8}\mathcal{S}_{1}\mathcal{S}_{2}\mathcal{S}_{3}}.
\]
However, the equality holds only when both sides vanish. To generalize this
relation, we introduce a constant $\tau_{s}^{\prime}$ and reformulate the
uncertainty relation as%
\[
\triangle H_{12}\triangle H_{23}\triangle H_{31}\geqslant\sqrt{\frac{\tau
_{s}^{\prime3}}{8}\mathcal{S}_{1}\mathcal{S}_{2}\mathcal{S}_{3}}.
\]

Our goal is to compute the value of $\tau_{s}^{\prime}$. From the above
inequality, we derive%
\[
\tau_{s}^{\prime}\leqslant\left(  \frac{8\left(  \triangle H_{12}\right)
^{2}\left(  \triangle H_{23}\right)  ^{2}\left(  \triangle H_{31}\right)
^{2}}{\mathcal{S}_{1}\mathcal{S}_{2}\mathcal{S}_{3}}\right)  ^{\frac{1}{3}}.
\]
Similarly, we examine the most general state given in Eq. (\ref{psi}), where
$\left(  \triangle H_{12}\right)  ^{2}$, $\left(  \triangle H_{23}\right)
^{2}$, $\left(  \triangle H_{31}\right)  ^{2}$, $\mathcal{S}_{1}$,
$\mathcal{S}_{2}$, and $\mathcal{S}_{3}$ are calculated in the Appendix. Based
on the results from the additive uncertainty relation, when $\left\vert
\psi\right\rangle =\left\vert \psi\right\rangle _{c_{1}}^{\prime}$ or
$\left\vert \psi\right\rangle =\left\vert \psi\right\rangle _{c_{2}}^{\prime}%
$, we find $\tau_{s\max}^{\prime}=\frac{2}{\sqrt{3}}$.

\section{Discussion and Conclusion}

This study establishes a novel method for identifying $3$-qubit $W$ states
through quantum uncertainty relations. By investigating an $XXZ$ Heisenberg
model composed of three spin-$\frac{1}{2}$ particles, we rigorously
demonstrate that when both the additive and multiplicative uncertainty
relations of the three observables $H_{12}$, $H_{23}$, and $H_{31}$
simultaneously saturate their equality conditions, the quantum state must be a
$W$ states. This discovery fundamentally reveals profound connections between
uncertainty relations and multipartite entanglement structures. Our work
provides the first direct demonstration linking uncertainty relations in $XXZ$
Heisenberg systems to $W$ states characterization, thereby revealing a
profound correspondence between entanglement structures and the critical
behavior of quantum uncertainties. Beyond its immediate application as a
theoretical tool for efficient $W$ states detection in experimental settings,
our approach establishes a versatile framework for investigating more complex
entanglement architectures, including $N$-qubit $W$ states ($N\geq4$) and
high-dimensional entangled systems.

\begin{acknowledgments}
		J.L.C. is supported by the Quantum Science and Technology-National Science and Technology Major Project (Grant No. 2024ZD0301000), and the National Natural Science Foundation of China (Grant No. 12275136).
      Z.J.L. is supported by the Nankai Zhide Foundation.
	
\end{acknowledgments}

\end{document}